\documentclass[aps,prl,pdflatex,twocolumn,notitlepage,superscriptaddress]{revtex4-2}
\pdfoutput=1

\usepackage[compat=1.0.0]{tikz-feynman}
\usepackage{amssymb}
\usepackage{multirow}
\usepackage{bm}
\usepackage{amsmath}
\usepackage{graphicx}
\usepackage{epstopdf}
\usepackage{subfigure}
\usepackage{natbib}
\usepackage{epsfig}
\usepackage{amsfonts}
\usepackage{mathrsfs}
\usepackage[toc,page,title,titletoc,header]{appendix}
\usepackage[colorlinks,linkcolor=blue,citecolor=blue,anchorcolor=blue]{hyperref}
\usepackage{braket}
\usepackage{dsfont,amsthm,amsbsy}
\usepackage{fancyhdr}
\usepackage{ulem}
\usepackage{bbm}
\usepackage{xcolor}

\def\Tr{{\rm Tr}}

\def\spinc{spin$_\mathbb{C}$ }

\begin{document}

\title{Single-component twisted $\mathbb{Z}_3$ orthogonal metal in an $e/3$-anyon fluid
}

\author{Zhaoyu Han}
\affiliation{Department of Physics, Harvard University, Cambridge MA 02138, USA}

\author{Ashvin Vishwanath}
\affiliation{Department of Physics, Harvard University, Cambridge MA 02138, USA}

\author{Eslam Khalaf}
\affiliation{Department of Physics, Harvard University, Cambridge MA 02138, USA}
\begin{abstract}

We propose an unconventional metallic phase emerging upon doping the $1/3$ Laughlin Fractional Chern insulator, which can serve as a parent state to anyon superconductivity with arbitrary chiral central charge. It is a twisted $\mathbb{Z}_3$
orthogonal metal: a state with vanishing electron quasiparticle weight but a single well-defined Fermi surface of emergent charge-$e/3$ fermions coupled to a Dijkgraaf-Witten twisted $\mathbb{Z}_3$ gauge field. It is manifestly symmetric under the projective action of translations on the three low-energy valleys and is fully gapped in the valley sector. Pairing this fractionalized Fermi surface then removes the topological order and produces a family of superconductors whose chiral central charge is directly set by the BdG band topology of the paired $e/3$ fermions, while the angular momentum of the physical order parameter is constrained to be a multiple of three. This scenario provides a route to superconducting states beyond anyon-superconductivity constructions based on $2e/3$ anyons. The metallic phase itself carries distinctive ``fractional Fermiology'' signatures: 
$e/3$ shot noise, anomalous quantum oscillations, and a $6\pi$ ac Josephson effect when it mediates the Josephson coupling between superconductors. We construct explicit wavefunction ansatz realizing the phase, and discuss energetics considerations that might favor this state over competing metallic states. We show that the construction extends to higher Laughlin states but not to Jain states.

\end{abstract}

\maketitle
The recent discovery of the fractional quantum anomalous hall (FQAH) effect in moir\'e materials~\cite{Cai2023SignaturesFQAHMoTe2,Park2023FractionallyQuantizedAnomalousHallMoTe2,Xu2023IntegerFractionalQAHMoTe2,Zeng2023ThermodynamicFCIMoTe2,xu2026signaturesunconventionalsuperconductivitynear} has brought a set of long-standing questions into sharp focus. In these systems, control over charge density allows us to inject carriers with fractional charge and anyonic statistics. The presence of a lattice potential generates a dispersion for these anyons~\cite{yan2025anyonDispersionAC,schleith2025anyonDispersionSphere,iyer2026anyonBlochBands, Goncalves}, both in the FQAH and in more general fractional Chern insulators (FCIs) stabilized in magnetic field~\cite{doi:10.1126/science.aan8458,xie2021fractional}, in contrast to their Landau level counterparts. These developments therefore motivate a fresh look at an old question: what happens when fractionalized insulators are doped with mobile carriers? Pioneering work decades ago pursued doped fractional states as an unconventional route to superconductivity (SC)~\cite{Laughlin1988SuperconductingFractionalStatistics,Laughlin1988HighTcFQH,FetterHannaLaughlin1989RPAFractionalStatisticsGas,ChenWilczekWittenHalperin1989AnyonSuperconductivity,Wilczek1990FractionalStatisticsAnyonSuperconductivity}. Recent developments have given this question renewed urgency and a growing body of theoretical and numerical work has followed  ~\cite{kcm5-hx56,6bgj-bfdn,doi:10.1073/pnas.2426680122,xs7g-dmn8,SciPostPhys.19.6.150,fan2026hiddenweakpairingsuperconductivitynoninteracting,lotric2026phasesitinerantanyonslaughlins,shi2026superconductivitynonfermiliquidmetals,pichler2026microscopic,ShiAnyonDelocalizationTransitions,shi2025nonabeliantopologicalsuperconductivitymelting,pichler2026microscopic,zhang2025holonmetalchargedensitywavechiral,wang2026topologicalsuperconductivityabelianfractional,shi2026charge4esuperconductorparafermionicvortices}. 

In particular, a variety of phases including SC, re-entrant integer quantum Hall (RIQH) and metallic states have been experimentally identified in the vicinity of the $2/3$ FQAH insulator~\cite{xu2026signaturesunconventionalsuperconductivitynear,Park2023FractionallyQuantizedAnomalousHallMoTe2}. A plausible route to SC emerges if one assumes that $2e/3$ anyons, rather than the minimal $e/3$ anyons, are the cheapest charge excitations of the $2/3$ FCI~\cite{Wilczek1990FractionalStatisticsAnyonSuperconductivity,6bgj-bfdn,ShiAnyonDelocalizationTransitions}. Indeed, recent numerical works noted a tendency of $e/3$ anyons to form $2e/3$ or higher charge bound states when the Coulomb interaction is suffiently screened~\cite{AnyonMoleculesJain, AnyonClustersYang, li2026boundstatesanyonsgeometric, WangZaletelAnyonMolecules,nddl-729x}. However, the appearance of metallic phases nearby, for instance, when the SC is destroyed by magnetic fields or temperature, motivates the search for anyon doping routes to SC with a parent metallic state. Additionally, recent numerics on simplified models of the twisted MoTe$_2$ FCI have also observed superconductivity upon doping the $2/3$ state, but with chiral central charge $c_-=-1/2$~\cite{wang2025chiralsuperconductivitynearfractional,zm39-dstj}, which differs from that predicted by the simplest $2e/3$ mechanism. These numerics come with important caveats, including small system sizes which makes it difficult to access the dilute doping regime, and it remains unclear whether the observed SC arises from an electron Fermi-surface instability or from an anyon mechanism. Nevertheless, taken together, these observations motivate a deeper look at what happens when the minimal charge $e/3$ anyons are doped.

Along these lines, a previous composite-fermion-based approach that pays attention to the multi-valley structure of the dispersion arising from fractionalized translation~\cite{kcm5-hx56} found that doping leads to a strongly interacting ``secondary composite fermion'' metallic phase that is ultimately unstable to a CDW coexisting with a small Fermi pocket. Such a state is consistent with the observed RIQH phase seen near $2/3$.  More recently, a different approach  introduced a ``color-valley locked'' metallic state on doping, argued to be ultimately unstable to a $c_-=-1/2$ superconductor ~\cite{fan2026hiddenweakpairingsuperconductivitynoninteracting, wang2026topologicalsuperconductivityabelianfractional, zhang2025holonmetalchargedensitywavechiral}.  Both proposals dope anyons with statistics $\theta=\pi/3$ and charge $e/3$; but differ in the choice of flux-attachment scheme to model them. In both cases, the nontrivial braiding statistics of taking anyons around each other produce gauge fluctuations, leading to the strongly coupled nature of the metallic states.

In this work, we develop a different approach. In this description, doping triggers an additional topological order which, when combined with that of the parent state, admits excitations carrying charge $e/3$ that are fermions rather than anyons. Upon doping, charge enters the system in the form of these particles. Because they braid trivially among themselves, gauge fluctuations play a minimal role and the resulting metallic state is of orthogonal-metal type~\cite{PhysRevB.86.045128}: it is a non-Fermi liquid, violates Luttinger's theorem and has vanishing overlap with the physical electron operator, but the emergent quasiparticles are weakly interacting and form a Fermi liquid of themselves. Crucially, this metal is valley-symmetric and valley-gapped: the three anyon valleys generated by fractionalized translations are all present in the parent theory, but no valley structure appears at the Fermi surface, and there is a  gap to valley excitations. A related $\mathbb{Z}_3$ orthogonal metal with three Fermi pockets was recently proposed in Ref.~\cite{shi2026superconductivitynonfermiliquidmetals}. Unlike the single-pocket state studied here, it retains gapless valley excitations and admits no valley-symmetric, fully gapped paired descendant.

In contrast to orthognal metals based on the $\mathbb{Z}_2$ toric-code topological order~\cite{PhysRevB.86.045128, gazit2017emergent}, pairing these fermions can produce SCs without leaving behind residual topological order. Furthermore, they can realize any possible value of chiral central charges, including $c_- = -1/2$. We explore this theory as a candidate description of the metallic states observed near the 2/3 FCI in twisted MoTe$_2$, and whether the superconductivity emerging at lower temperatures could be a pairing instability of the orthogonal Fermi surface.
We point out experimental signatures that could distinguish this fractionalized metal from the alternatives, and briefly discuss candidate wavefunctions and the energetics considerations that might favor this state.

{\bf The model and the emergent valley symmetry. } We start from a composite-boson~\cite{PhysRevLett.62.82} description of charge-$e/3$ anyons near the $\nu=1/3$ Laughlin state on a lattice:
\begin{align}
\mathcal{L} = \mathcal{L}[\phi;\alpha] - \frac{3}{4\pi} \alpha\mathrm{d}\alpha  + \frac{1}{2\pi} \alpha\mathrm{d}(A+\frac{3}{2}\omega) .
\label{eq:cb}
\end{align}
Here $\phi$ is a bosonic field sourcing the doped anyons, $A$ is the electromagnetic probe field, and $\omega$ is the $SO(2)$ spin connection of the underlying manifold, or its crystalline analogue, which tracks orbital angular-momentum response. The mixed term $\alpha \mathrm{d}\omega$ is the Wen-Zee coupling: its coefficient fixes the orbital spin of the anyons~\cite{WenZeeShift,PhysRevB.90.115139}. 

We assume that the lattice system is at filling $\nu=1/3+\delta$ per crystal unit cell. The average flux density of $\alpha$ is therefore $
\frac{\langle \nabla\times {\bf \alpha}\rangle}{2\pi}=\frac{1}{3}+\delta$, and the density of doped anyons is $\rho_\phi=3\delta$. The fractional background flux $1/3$ per unit cell enlarges the magnetic unit cell by a factor of three and produces three degenerate valleys in the reduced magnetic Brillouin zone for the doped anyons. We decompose the low-energy boson field as $\phi\rightarrow \phi_v$, $v=0,1,2$, where the commensurate $1/3$ flux is absorbed into the valley structure. There is an  emanant~\cite{SciPostPhys.15.2.051} $\mathbb{Z}_3\times \mathbb{Z}_3$ valley symmetry descending from microscopic translation symmetries $T_{1,2}$: in a chosen gauge detailed in End Matter, $\phi_v$ fields transform as $T_1: \phi_v \mapsto  \phi_{v+1}$ and $T_2: \phi_v \mapsto \omega^{v+1} \phi_{v+1}$  where $\omega=e^{\mathrm{i} 2\pi/3}$~\cite{PhysRevB.72.134502}.  Assuming $C_3$ or $C_6$ rotational symmetry, the general form of the long-wavelength ($\mathcal{O}(\delta^2)$) Lagrangian density is
\begin{align}
\mathcal{L}[\phi_v;\alpha] = &\sum_v \phi^*_v\left[ \mathrm{i}\partial_0+\alpha_0 - \frac{1}{2m}\left(-\mathrm{i}{\bf \nabla}-{\bf \alpha}\right)^2 \right]  \phi_v +\mathcal{L}_{\text{int}} \nonumber\\
\mathcal{L}_{\text{int}}=&V_0\rho^2 + 
    V_1\left|\eta
    \right|^2,
\label{eq:valleyphi}
\end{align}
where $m$ is the anyon mass, and $\rho \equiv \sum_v |\phi_v|^2$ is the density, and $\eta\equiv \sum_v\omega^v |\phi_v|^2$ is a `valley moment' density which doesn't commute with $T_1,T_2$ but preserves $T_1 T_2^{-1}$. In End Matter, we prove that these interactions terms are the only quartic terms allowed by the spatial symmetries, and derive explicit expression for $V_{0,1}$ in terms of the microscopic anyon density-density interaction. In principle, these parameters can be fully extracted from microscopic calculations, e.g. in the Laughlin state in Aharonov-Casher bands~\cite{yan2025anyonDispersionAC, li2026boundstatesanyonsgeometric}. To this order of expansion, Eq.~\eqref{eq:valleyphi} has an emergent valley symmetry $G_\text{valley} = \frac{U(1)^3\rtimes\mathbb{S}_3}{U(1)}$, which is further enlarged to $PSU(3)$ at the special point $V_1=0$. 

{\bf Single-component $\mathbb{Z}_3$ orthogonal metal. } We now introduce the parton decomposition
\begin{align}\label{eq:parton}
\phi_v = f \cdot  g_v ,
\end{align}
implemented by
\begin{align}
\mathcal{L}[\phi_v;\alpha] = \mathcal{L}[f;a] + \mathcal{L}[g;\alpha-a] ,
\end{align}
where $a$ is an emergent dynamical gauge field~\footnote{Strictly, $a$ is a \spinc connection: its elementary sources are fermions, and a $\pi$ holonomy on a non-contractible cycle changes the associated fermion boundary condition. This distinction does not change the response formulas used below, but it is useful for tracking the spin-charge relation and the statistics of the resulting anyons~\cite{SEIBERG2016395,10.1093/ptep/ptw083,SENTHIL20191,han2023fractional}.}. We choose the ansatz with zero average $a$ flux, so that the $f$ fermions see no flux and can form a Fermi liquid. Each species of $g_v$ fermion has density $\delta$ and sees the incommensurate flux $\frac{\langle \nabla\times {\bf \alpha}\rangle}{2\pi}=\delta$, thus is at effective filling fraction $\nu=1$ and can be put in an integer quantum Hall (IQH) state. This ansatz manifestly preserves the valley symmetry. We show in End Matter that this construction readily generalizes to other Laughlin states, but not to general Jain states.

Integrating out the gapped $g_v$ fermions gives
\begin{align}
\mathcal{L} =& \mathcal{L}[f;a] + 3\text{CS}[a,g] -  \frac{3}{2\pi} \alpha\mathrm{d}a +\frac{1}{2\pi} A\mathrm{d}\alpha  +\frac{1}{2\pi}  \frac{3}{2}a\mathrm{d}\omega .
\label{eq:om}
\end{align}
Here we use the shorthand notation $\text{CS}[a,g]\equiv a\mathrm{d}a/(4\pi)+\Omega_g$ for the response of a filled Chern band, including the gravitational Chern-Simons term $\Omega_g$ that contributes chiral central charge $c_-=1$. Since the $f$ fermions see no net magnetic field, they may form a Fermi liquid with volume fixed by density $3\delta$. Its detailed dispersion is non-universal and depends on microscopic band geometry and interactions, but there is no symmetry-enforced valley multiplicity in the $f$ sector and generically there should be a single Fermi pocket without additional point-group symmetry. The $\alpha$ field effectively Higgses $a$ to a Dijkgraaf-Witten-twisted $\mathbb{Z}_3$ gauge field. The resulting phase is thus a single-pocket $\mathbb{Z}_3$ OM.

The stability of this phase is plausible already at the mean-field level. Although the long-distance gauge structure is discrete, the $\mathbb{Z}_3$ gauge theory can be viewed as the Higgs descendant of a parent $U(1)$ gauge theory. The massive remnant of the gauge fluctuation then mediates a short-ranged density-density interaction among the $f$ fermions. With the conventional sign of the gauge-field stiffness, this interaction is repulsive. This is analogous to the discussion on the standard $\mathbb{Z}_2$ OM~\cite{PhysRevB.86.045128,PhysRevB.62.7850,SenthilFisher2000Fractionalization}. The state should therefore be viewed as a legitimate compressible fixed point, with the usual caveat that sufficiently attractive pairing channels can destabilize it at the lowest temperature, as we will analyze below.

{\bf Fractionalized fermiology. } The OM is a minimal non-Fermi liquid~\cite{PhysRevB.86.045128}: it has metallic thermodynamics and transport from a hidden Fermi surface, near which there are long lived quasiparticle excitations but the physical electron has no overlap with these excitations, which is easily seen from the fact that they carry fractional charge. To see this, we examine the equations of motion of Eq.~\ref{eq:om} by varying $a$ and $\alpha$:
\begin{align}
(J_f)^\mu  &= \frac{3}{2\pi}\epsilon^{\mu\nu\lambda}\partial_\nu(\alpha_\lambda-a_\lambda),
\label{eq:jf}\\
3 \epsilon^{\mu\nu\lambda}\partial_\nu a_\lambda &= \epsilon^{\mu\nu\lambda}\partial_\nu A_\lambda .
\label{eq:constraint}
\end{align}
On the other hand, the physical current is
\begin{align}
J^\mu = \frac{1}{2\pi}\epsilon^{\mu\nu\lambda}\partial_\nu \alpha_\lambda
= \frac{1}{3}\frac{1}{2\pi}\epsilon^{\mu\nu\lambda}\partial_\nu A_\lambda + \frac{1}{3}(J_f)^\mu .
\label{eq:physcurrent}
\end{align}
Thus a quasiparticle excitation at the $f$ Fermi surface carries charge $e/3$ above the FCI background, and the elementary charge inferred from shot noise should be $e/3$.

The same constraint produces anomalous quantum oscillations. In an external magnetic field $B=\nabla\times\bf A$, the $f$ fermions see effective field $\nabla\times {\bf a} = B/3$. The Fermi volume is $\rho_f=3\delta$. For a spinless two-dimensional Fermi surface, Onsager quantization gives an oscillation frequency proportional to the Fermi surface area divided by the effective magnetic field coupled to the fermions. Equivalently, because the density is enlarged by a factor of three while the effective magnetic field is reduced by a factor of three, the oscillation frequency in $1/B$ is
\begin{align}
\tilde F = \frac{9|\delta|}{2\pi},
\end{align}
up to conventional factors of $e$ and $\hbar$ that are set to unity here. This is nine times the frequency of a conventional spinless Fermi liquid at the same physical excess charge density $\delta$.

The transport signatures are also fractionalized. Eqs.~\eqref{eq:constraint}\&\eqref{eq:physcurrent} relate the physical conductivity to that of $f$:
\begin{align}
\sigma^\text{phys} = \left(\frac{\sigma^f}{9}+\sigma^{1/3}\right),
\end{align}
where $\sigma^{1/3}$ is conductivity tensor of the parent $\nu=1/3$ Laughlin state. If the $f$ fermions have Drude  longitudinal conductivity $\sigma^f_{xx} = \frac{\rho_f \tau}{m_f}$, then
\begin{align}
\sigma^\text{phys}_{xx}=\frac{\sigma^f_{xx}}{9}=\frac{|\delta|\tau}{3m_f}.
\end{align}
Thus, for the same effective mass, scattering time $\tau$ and charge density $|\delta|$, the longitudinal conductivity of this state is one third of that of a conventional spinless charge-$e$ Fermi liquid.

\begin{figure}[t]
    \centering
    \includegraphics[width=\linewidth]{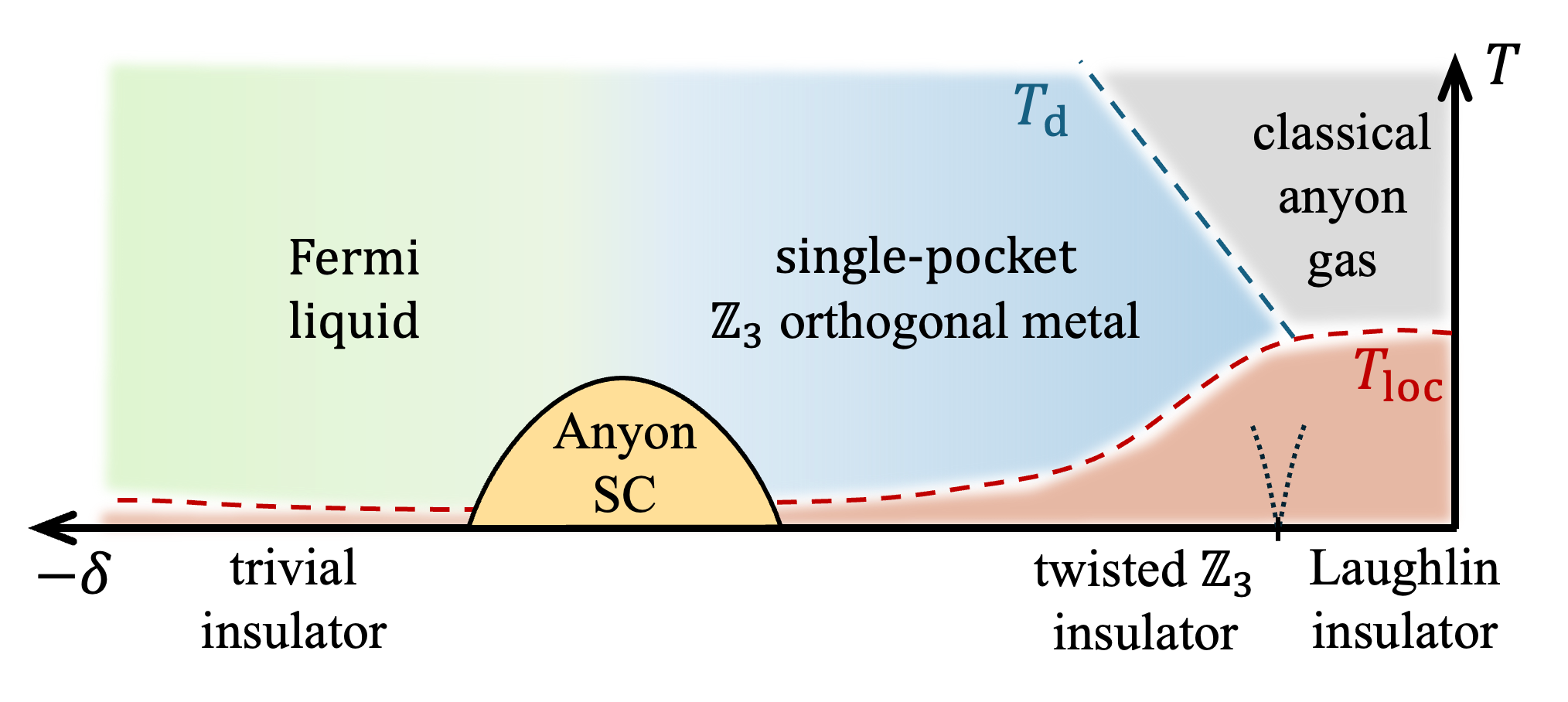}
    \caption{A possible phase diagram as a function of doping charge density $\delta$ and temperature $T$. $T_\text{d}$ and $T_\text{loc}$ are, respectively, the quantum degeneracy and localization scales of the anyons. An alternative possibility (not shown) is that the  $\mathbb{Z}_3$ orthogonal metal and the Fermi liquid (and correspondingly their zero temperature insulating phases) are separated by a first order transition.}
    \label{fig:schematic}
\end{figure}

{\bf The effects of disorder. } Since the low-energy charge dynamics are governed by a Fermi gas of $f$ fermions, arbitrarily weak disorder ultimately localizes the $f$ sector at sufficiently low temperature. Below a characteristic scale $T_{\rm loc}$, the $f$ fermions are effectively localized, and the zero-temperature phase is a fractionalized Anderson insulator within the mean field picture. The remaining intrinsic topological order is a Dijkgraaf-Witten-twisted $\mathbb{Z}_3$ toric code. This insulating phase has the same quantized charge response as the $\nu=1/3$ Laughlin state, but has a nine-fold topological degeneracy on a torus for any fixed disorder realization. For scattering time $\tau$ and effective mass $m$, the localization scale can be estimated from two-dimensional unitary-class scaling as
\begin{align}
    T_{\rm loc}\sim \tau^{-1}\exp\left[-{\rm const.}\left(\frac{\tau\delta}{m}\right)^2\right],
\end{align}
up to nonuniversal constants~\cite{RevModPhys.80.1355}. Thus, $T_{\rm loc}$ drops rapidly with increasing $|\delta|$, and the OM survives over a wide intermediate window $T_{\rm loc}<T<T_{\rm d}$, where $T_{\rm d}\sim |\delta|/m$ is the quantum degeneracy scale of the anyons. Above $T_{\rm d}$, the system crosses over to an incoherent classical gas.

Disorder also affects the $g_v$ sectors. In the clean construction, each $g_v$ species forms a $\nu=1$ IQH state. With disorder, this IQH state is reached through a plateau transition as the dopant density is increased. The relevant cyclotron scale grows as $\omega\sim |\delta|/m$, and can drive each $g_v$ sector from a trivial localized regime with Chern number zero to the clean-limit Chern number $C=1$ when $\omega \sim 1/ \tau$, which leads to a critical density $\delta_c \sim m/\tau$. This transition is neutral: it changes the emergent topological order, from that of the parent Laughlin state to the Dijkgraaf-Witten-twisted $\mathbb{Z}_3$ gauge theory, which we call `twisted $\mathbb{Z}_3$ insulator' and is essentially the Laughlin state stacking with a neutral bosonic $\mathbb{Z}_3$ topological order; this transition changes the chiral central charge $c_-$ from $1$ to $3$, but is invisible in the physical electrical conductivity at low temperature. We mention that the structure of this neutral plateau transition depends on valley mixing. If the disorder potential is smooth on the lattice scale, the emergent valley symmetry remains approximately intact and the three $g_v$ sectors experience nearly identical disorder; the transition can then occur in a single step. More generally, the three valley sectors need not transition simultaneously, and the transition can split into multiple  steps. This is similar to the discussions of FCI-SC transition driven by $2e/3$ anyon plateau transition~\cite{6bgj-bfdn,ShiAnyonDelocalizationTransitions}. 

At large anyon doping $|\delta|$, the effective disorder strength is reduced such that SC can emerge if there were attraction among $f$ particles. Since the resulting SC is not fractionalized as we will show below, it can connect smoothly to a SC that emerges from a Fermi liquid when $|\delta|$ is further increased. More generally, as $|\delta|$ increases, the fractionalized excitations are no longer in the dilute limit, presumably triggering phase transitions to other states. Putting together, we summarize the resulting schematic $\delta$-$T$ phase diagram in Fig.~\ref{fig:schematic}.

{\bf Pairing descendants. } We next consider paired states of the $f$ Fermi surface. Different pairing channels are characterized by the Majorana Chern number $C_{\rm maj}$ of the occupied BdG bands and by the angular momentum $l$ of the $f$ pairing order. Because the valley dependence is only carried by $g_v$, pairing $f$ does not break any valley symmetry. 

The condensation of $f$ pairs $\Delta_f \sim ff$, which are bosonic, charge-$2e/3$ anyons, confines all the anyons in the parent topological order and leads to a charge-$2e$ SC. This process changes the total chiral central charge only by $C_\text{maj}/2$, leading to a non-fractionalized SC with
\begin{align}
c_-=C_\text{maj}/2 +3.
\end{align}
The physical gauge-invariant SC order parameter is $\Delta_\text{phys}\sim \Delta_f^3 \mathcal{M}_\alpha^2$ where $\mathcal{M}_\alpha$ is the $2\pi$ monopole operator of $\alpha$, suggesting that the angular momentum of the physical order is 
\begin{align}
L=3l
\end{align}
modulo crystalline symmetry. We show in End Matter a more careful field theoretical derivation of these results. 

In particular, a $f-if$ weak pairing of $f$ with $C_{\rm maj}=-3$ would, after particle-hole conjugation, lead to a SC with a $c_-$ and $L$ consistent with the observed SC in numerics near the $2/3$ FCI state on a hexagonal lattice~\cite{wang2025chiralsuperconductivitynearfractional,zm39-dstj}. An odd-angular-momentum strong pairing with $C_{\rm maj}= 0$ leads to a SC with $c_-$ and $L$ consistent with that obtained from the Laughlin's $2e/3$ anyon SC mechanism~\cite{kcm5-hx56,6bgj-bfdn}.

These results sharpen the role of the single-pocket OM: Pairing can generate an infinite sequence of Abelian and non-Abelian SCs, with integer or half-integer $c_-$ determined by the BdG topology of the hidden Fermi surface. In this view, the present single-pocket OM is therefore not merely another compressible anyon fluid, but a symmetric parent from which different anyon SCs descend.

Since the superconducting descendants can be understood as paired states of the $f$ fermions, a junction in which two superconducting regions are separated by the OM can be governed by coherent tunneling of an $f$ pair through the metallic region. An $f$ pair carries physical charge $2e/3$, rather than the charge $2e$ of an ordinary Cooper pair. The leading fractional Josephson coupling then has a $6\pi$ periodic current-phase relation. Equivalently, under a DC voltage bias $V$, the AC Josephson frequency is reduced to
$\omega_J=\frac{2V}{3}$.

{\bf Wave-function construction. } We now present a microscopic wave-function ansatz for the single-pocket OM formed by quasi-holes of an $1/m$ Laughlin state in Aharonov-Casher bands~\cite{Aharonov1979GroundStateSpinHalf}, which are lattice analogues of lowest Landau level ideal for hosting Laughlin FCI in the presence of short-ranged repulsions~\cite{PhysRevB.108.205144,PhysRevB.31.5280}. The starting point is the symmetric, anti-holomorphic anyon wavefunction function $\Psi(\bar{\xi})$ of $M$ quasi-hole coordinates $\{\xi_1 \dots \xi_M\}$. In the geometric quantization approach~\cite{yan2025anyonDispersionAC,li2026boundstatesanyonsgeometric}, the Berry phase of an anyon trajectory can be captured by a K\"ahler potential
\begin{align}
    {\cal K}(\bar \xi,\xi)
    =
    \ln \langle \psi_{\xi_1,\ldots,\xi_M}|\psi_{\xi_1,\ldots,\xi_M}\rangle .
\end{align}
Given this K\"ahler potential and the microscopic crystal potential, the single-particle anyon band structure can be obtained directly~\cite{yan2025anyonDispersionAC}. 
The band-structure data are the only microscopic input needed to define the variational ansatz for the single-pocket OM and its SC descendants.

Here we give the wavefunction ansatz corresponding to the parton construction in Eq.~\eqref{eq:om}. Let $\psi_{\bf k}(\bar \xi)$ be a single-anyon Bloch wave function, and suppose that its lowest band has valley minima at momenta $\bf K_\alpha$, $\alpha=1,\ldots,m$, then the ansatz reads:
\begin{align}
    \Psi(\{\bar \xi\})
    =& f(\{\bar{\xi}\}) g(\{\bar{\xi}\}) \\
  g(\{\bar{\xi}\})=& \det\left[u_i(\bar{\xi}_j)\right] 
    \label{eq:valley_expansion_main}
\end{align}
where $i = (\alpha, r) $ is a composite index with ranges $\alpha=0, \dots, q-1$ and $r = 0, \dots, M/3-1$, and
\begin{align}
    u_i(\bar{\xi}_j) = \partial^r_{\bar{q}} \left. \psi_{{\bf K}_{\alpha}+{\bf q}}(\bar{\xi}_j) \right|_{\bf q=0}
\end{align}
Here $g$ is essentially a fully antisymmetric wavefunction of a fermion Hoftadter problem defined by a dilute filling $3|\delta|$ and  $1/3-|\delta|$ average flux per unit cell. Then, the effective K\"ahler potential for the wavefunction $f$ is obtained by
\begin{align}
    \widetilde{\cal K}
    =
    {\cal K}
     - \ln|g|^2 ,
    \label{eq:Ktilde_main}
\end{align}
We will show in an upcoming paper that the background and statistical fluxes in Eq.~\eqref{eq:Ktilde_main} are all exactly canceled, so the fermionic factor $f$ defines a fermionic problem seeing no zero net magnetic field and no residual statistical flux at long wavelengths. Therefore, one may further consider the wavefunctions where $f$ describes a Fermi liquid or its paired descendant.

{\bf Competing metallic states. } The single-pocket OM is not the only possible metallic phase of the multi-valley anyon problem in Eq.~\ref{eq:valleyphi}. Here we discuss three natural competitors.

The first is the triple-pocket OM discussed in Ref.~\cite{shi2026superconductivitynonfermiliquidmetals}. This state is obtained by assigning all the internal flux to the $f$ sector, so that the $f$ partons form a $\nu=3$ IQH state. Although closely related, it is physically distinct from the single-pocket OM. In the single-pocket OM, the valley-carrying excitations are gapped, whereas in the triple-pocket OM they remain gapless because of the valley-resolved Fermi surfaces. The triple-pocket OM is therefore expected to exhibit stronger low-energy fluctuations of period-three density-wave orders associated with inter-valley particle-hole bilinears. Upon pairing, the state cannot gap all Fermi surfaces while fully preserving the translation symmetry, let alone the larger emergent valley symmetry.

The second possibility is another triple-pocket metal, considered in Ref.~\cite{fan2026hiddenweakpairingsuperconductivitynoninteracting}, in which each valley field is fractionalized independently as $\phi_v=f_v g_v$. When each $g_v$ is placed in a $\nu=1$ IQH state, the resulting phase contains three Fermi pockets carrying different emergent gauge charges. We therefore refer to it as a triple-pocket ``color-valley-locked" metal. This construction does not manifestly accommodate the approximate valley $PSU(3)$ symmetry near $V_1=0$. However it has the virtue of realizing the mutual  statistics of anyons in pockets $I,\,J$ with the minimal flux attachment consistent with vanishing net flux $\Phi_{IJ}/2\pi=-\delta_{IJ} +1/3$.

Finally, there are metallic states that explicitly break the valley symmetry. For example, suppose all anyons populate a single valley, say $v=1$. Placing $f$ in a $\nu=1$ IQH state then puts the polarized $g_{1}$ partons at effective filling $\nu_f=-3/2$. A natural construction is for $f$ to fill the lowest Landau level and form a composite Fermi liquid in the first Landau level, realizing a valley polarized version of the ``secondary composite Fermi liquid'' considered in Ref.~\cite{kcm5-hx56}. This construction ultimately produces no intrinsic fractionalization, and the valley polarization should be viewed as producing a period-three charge density wave (CDW). We henceforth refer to the resulting phase as a CDW metal. 

Although a microscopic calculation is needed to determine the true ground state, we can nevertheless conjecture regimes in which the different metallic states are natural candidates. When the inverse mass $1/m$ is small compared with  $V_{0,1}$, it is reasonable to expect the competition to be controlled primarily by the ratio $\frac{V_1}{V_0}$. We note that, unlike $\rho^2$ term which is insensitive to valley structure, $|\eta|^2$ interaction effectively measures the square of the valley moment density $\eta$, and is thus sensitive to the fluctuation of valley excitations violating the valley symmetries. Among the four states considered above, the single-pocket OM is the only one that retains a gap to valley excitations. Therefore, it can be expected that $\langle|\Phi|^2\rangle$ is smallest in this state and thus this state can be stabilized by a large positive $V_1$. On the other hand, we note that the interaction can be re-written as
\begin{align}
    {\cal L}_{\mathrm{int}}
    =
    (V_0+V_1)\sum_{v}
    |\phi_v|^4
    + (2V_0-
    V_1)\sum_{v}
    |\phi_v|^2|\phi_{v+1}|^2
\end{align}
such that $(V_0+V_1)$ and $(2V_0- V_1)$ measure the intra- and inter-valley repulsions. When the inter-valley repulsion is large, we expect that the valley polarized state or other correlated state with suppressed inter-valley coincidence to be favored. 
Finally, upon including the effects of anyon dispersion, the color-valley locked metal may be favored,  since the distributing the dopants among three pockets reduces the Fermi momentum of the partons.

Experimentally, the fractionalized metallic states discussed above share similar features such as $e/3$ shot noise and reduced longitudinal conductivity, particularly the three pocket OM whose excitations are also weakly interacting fermionic quasiparticles with $e/3$ charge. Furthermore, the triple-pocket OM and triple-pocket color-valley locked metal could also mediate fractionalized Josephson effects between appropriate anyon SCs. One distinguishing feature of the single-pocket OM is the largest quantum oscillation frequency among all candidates, which is a factor of three larger than the other triple-pocket candidates and a factor of nine larger than a conventional state such as a CDW metal with a single Fermi pocket.

A useful next step would be a variational comparison in a concrete microscopic model to resolve the phase diagram of the competing metallic states. In terms of the effective anyon coordinates, one can also perform exact diagonalization with a relatively large number of anyons, providing a complementary numerical route. Finally, in a single-pocket OM, the physics may be dominated by the Fermi surface, for which mature analytical and numerical techniques are available. It would therefore be useful to adapt these methods to determine which pairing channel is favored once realistic band geometry and interactions are included.

{\bf Note added. } Near the completion of this work, we became aware of a related work~\cite{senthil2026fractionalizedmetalsdopedanyons} that independently develops the same construction. Our results are consistent in the aspects where they overlap.

{\bf Acknowledgement. } We thank Zhengyan Darius Shi, T. Senthil, Zijian Wang and Yahui Zhang for inspiring discussions. We particularly thank Qingchen Li for pointing out a simpler presentation of the wavefunction ansatz. Z.~H. and A.~V. are supported by the Simons Investigator award, the
Simons Collaboration on Ultra-Quantum Matter, which
is a grant from the Simons Foundation (651440, A.~V.). E.~K. is supported by NSF CAREER grant DMR award No. 2441781. 

\bibliography{ref}

\newpage
\appendix

\onecolumngrid
\begin{center}
    {\large \textbf{End Matter}}
\end{center}
\twocolumngrid

{\bf Generalizations to other fractions. } The construction generalizes directly to the Laughlin sequence at filling $\nu=1/m$, with odd $m$. The charge-$e/m$ anyon sees a fractional background flux $1/m$ per unit cell, producing $m$ valleys related by magnetic translations. Repeating the same parton construction gives a valley-symmetric $\mathbb{Z}_m$ OM with a single hidden Fermi surface, and its pairing descendants have the same structure as above: the chiral central charge is controlled by the BdG topology of the hidden Fermi surface, while the angular momentum of the physical superconducting order parameter is constrained to be a multiple of $m$. All the phenomenology of the state generalizes naturally.

However, this construction of a fractionalized metal does not naturally extend to generic Jain states. To see this, remember that the key step above is that the $g_v$ partons can fill an IQH state in each valley. This requires that the anyon self-statistical flux number to be $\theta/\pi = \pm 1/N_v$ mod $2$, where $N_v$ is the number of valleys determined by the background flux seen by the anyons. [More generally, the anyon self-statistical flux number can be $1/(kN_v)$ mod $2$ with $k>1$, which corresponds to putting each $g_v$ to $\nu=k$ IQH. However, for Jain states this can never happen.] This matching is special to Laughlin FCIs. 

To see the obstruction for the positive Jain sequence $\nu=p/(2p+1)$ with $p>0$ (the negative sequence can be obtained through charge conjugation), we first note that the background flux number per unit cell seen by the charge-$qe/(2p+1)$ anyon is $q/(2p+1)$ so that $N_v =\text{gcd}(q,2p+1)$, and the self-statistical flux number is 
$\theta/\pi = q^2(2p-1)/(2p+1)$ mod $2$. In order for $\theta/\pi = \pm 1/N_v$ mod $2$ to hold true, one can show that $p,q$ must be chosen in the following way: choose an odd $N>1$ and an integer $c$ that is less than and coprime with $N$, then solve an odd $d$ such that $2dc^2 = \pm 1$ mod $N$. For each choice of $(N,a)$, there are infinitely many choices of $d$, each giving a satisfying solution: $2p+1 = dN$, $q=dc$. However, these solutions are rather sparse, and the simplest example beyond the $e/3$ anyon at $\nu=1/3$ is the $5e/7$ anyon at $\nu=3/7$.

{\bf Projective symmetry and effective valley interaction. } In the Landau gauge that diagonalizes $T_1^{-1}T_2$, the primitive translations act projectively as
\begin{align}
    T_1:\phi_\ell&\mapsto\phi_{\ell+1},
    &
    T_2:\phi_\ell&\mapsto\omega^{\ell+1}\phi_{\ell+1},
    \label{eq:endmatter_translations}
\end{align}
with indices understood modulo three, so that
\(T_1T_2=\omega T_2T_1\). Correspondingly a convenient representatives of the point-group actions are
\begin{align}
    U_{C_3}
    &=
    \operatorname{diag}(1,\omega,\omega),
    &
    U_{C_6}
    &=
    \begin{pmatrix}
        1&0&0\\
        0&0&1\\
        0&\omega&0
    \end{pmatrix}.
    \label{eq:endmatter_rotations}
\end{align}

To connect the effective interaction to a microscopic translationally invariant density interaction, which is periodic in crystal Brillioun zone (CBZ):
\begin{align}
    H_{\rm int}
    =
    \frac12\sum_{{\bf q}\in{\rm CBZ}}
    V({\bf q})\rho({\bf q})\rho(-{\bf q}),
    \qquad
    V({\bf q})=V(-{\bf q})\in\mathbb R,
\end{align}
we organize the gauge-neutral valley bilinears according to their crystalline momenta:
\begin{align}
    N&=\sum_{\ell}|\phi_\ell|^2,
    \nonumber\\
    D&=
    |\phi_0|^2+\omega|\phi_1|^2+\omega^2|\phi_2|^2,
    \nonumber\\
    O_r&=
    \sum_{\ell=0}^{2}
    \omega^{r\ell}\phi_\ell^\dagger\phi_{\ell+1},
    \qquad r=0,1,2.
    \label{eq:endmatter_bilinears}
\end{align}
Here \(D\) is associated with
\begin{align}
    {\bf Q}_B=\frac{{\bf G}_2-{\bf G}_1}{3},
\end{align}
whereas the \(O_r\) occupy the $C_6$ symmetric momentum star
\begin{align}
    \left\{
        \pm\frac{{\bf G}_1}{3},
        \pm\frac{{\bf G}_2}{3},
        \pm\frac{{\bf G}_3}{3}
    \right\}
\end{align}
which we call ${\bf Q}_{A,r}$. Projecting the microscopic density onto these harmonics gives schematically
\begin{align}
    \rho({\bf 0})&\simeq F_0N,
    &
    \rho({\bf Q}_B)&\simeq F_BD,
    &
    \rho({\bf Q}_{A,r})&\simeq F_{A,r}O_r,
\end{align}
where the form factors \(F_0,F_B,F_{A,r}\) are microscopic density matrix elements between the magnetic Bloch states. For \(C_3\) or \(C_6\) symmetry,
\(|F_{A,r}|=|F_A|\), and the projected interaction is
\begin{align}
    {\cal L}_{\rm int}
    =
    \frac{g_0}{2}N^2
    +
    g_A\sum_r|O_r|^2
    +
    g_B|D|^2,
\end{align}
with
\begin{align}
    g_0&=V({\bf 0})|F_0|^2,
    &
    g_A&=V({\bf Q}_A)|F_A|^2,
    &
    g_B&=V({\bf Q}_B)|F_B|^2.
\end{align}
Using
\begin{align}
    \sum_r|O_r|^2=N^2-|D|^2,
\end{align}
this reduces to
\begin{align}
    {\cal L}_{\rm int}
    =
    V_0
    \left(\sum_\ell|\phi_\ell|^2\right)^2
    +
    V_1
    \left|
        |\phi_0|^2+\omega|\phi_1|^2+\omega^2|\phi_2|^2
    \right|^2,
    \label{eq:endmatter_effective_interaction}
\end{align}
where
\begin{align}
    V_0
    &=
    \frac12V({\bf 0})|F_0|^2
    +
    V({\bf Q}_A)|F_A|^2,
    \nonumber\\
    V_1
    &=
    V({\bf Q}_B)|F_B|^2
    -
    V({\bf Q}_A)|F_A|^2.
    \label{eq:endmatter_V0V1}
\end{align}

For generic \(V_1\neq0\), the quartic theory is invariant under independent phase rotations and permutations of the three modes,
\begin{align}
    \phi_\ell\mapsto e^{i\theta_\ell}\phi_{\sigma(\ell)},
    \qquad
    \sigma\in S_3,
\end{align}
giving the emergent valley symmetry
\begin{align}
    G_{\rm valley}
    =
    \frac{U(1)^3\rtimes S_3}{U(1)_{\rm gauge}}
    \simeq
    U(1)^2\rtimes S_3.
\end{align}
At \(V_1=0\), this enlarges to
\begin{align}
    G_{\rm valley}=U(3)/U(1)\simeq PSU(3).
\end{align}

{\bf Field theory description to the pairing descendants. } For even Majorana Chern number, $C_{\rm maj}=2N$, the paired $f$ sector is represented by
\begin{align}
\mathcal{L}[f;a]  \rightarrow \frac{2}{2\pi}\beta\mathrm{d}(a+l\omega) + N \text{CS}[a,g] .
\end{align}
Substitution into 
\begin{align}
\mathcal{L} =& \mathcal{L}[f;a] + 3\text{CS}[a,g] -  \frac{3}{2\pi} \alpha\mathrm{d}a +\frac{1}{2\pi} A\mathrm{d}\alpha  +\frac{1}{2\pi}  \frac{3}{2}a\mathrm{d}\omega .
\label{eq:om}
\end{align}
gives
\begin{align}
\mathcal{L} = &\frac{2}{2\pi}\beta\mathrm{d}a  +(N+3)\text{CS}[a,g] -  \frac{3}{2\pi} \alpha\mathrm{d}a \nonumber\\
& +\frac{1}{2\pi} A\mathrm{d}\alpha   +\frac{1}{2\pi}  (l\beta +\frac{3}{2}a)\mathrm{d}\omega .
\end{align}
After the unimodular field redefinitions $\beta\rightarrow\beta+\alpha$ and then $\alpha\rightarrow\alpha+2\beta$, integration over $\alpha$ yields
\begin{align}
\mathcal{L}=\frac{1}{2\pi}\beta\mathrm{d}(2A+3l\omega) +(N+3)\text{CS}[A,g]+\cdots .
\end{align}
This describes an Abelian anyon superconductor with $c_-= N+3$, and the physical order parameter, the $4\pi$ monopole operator of $\beta$, carries angular momentum $3l$.

For odd Majorana Chern number, $C_{\rm maj}=2N+1$, a convenient description uses the $U(2)_{2,0}$ TQFT,
\begin{align}
\mathcal{L}[f;a] \rightarrow& - \frac{2}{4\pi} \Tr\left(\beta \mathrm{d} \beta+\frac{2}{3}\beta^3\right) + \frac{1}{4\pi} \Tr\beta\mathrm{d}\Tr\beta \nonumber\\
&+\frac{1}{2\pi} \Tr \beta \mathrm{d} \left(a +\frac{l}{2}\omega\right) +(N-1)\text{CS}[a,g] .
\end{align}
A parallel manipulation gives
\begin{align}
\mathcal{L} = & -\frac{2}{4\pi} \Tr\left(\beta \mathrm{d} \beta+\frac{2}{3}\beta^3\right) + \frac{1}{4\pi} \Tr\beta\mathrm{d}\Tr\beta \nonumber\\
& +\frac{1}{2\pi}\Tr\beta\mathrm{d}\left(A+\frac{3l}{2}\omega\right) +(N+2) \text{CS}[A,g] .
\end{align}
This is a non-Abelian anyon SC with $c_-= N+2+3/2$, and the physical order parameter, the $4\pi$ monopole operator of $\Tr\beta$, carries angular momentum $3l$.

\end{document}